\documentclass[twocolumn, switch]{article} 

\usepackage{preprint}

\usepackage{amsmath, amsthm, amssymb, amsfonts}

\usepackage[numbers,square]{natbib}
\usepackage[utf8]{inputenc}	
\usepackage[T1]{fontenc}	
\usepackage{xcolor}		
\usepackage[colorlinks = true,
            linkcolor = purple,
            urlcolor  = blue,
            citecolor = cyan,
            anchorcolor = black]{hyperref}	
\usepackage{booktabs} 		
\usepackage{nicefrac}		
\usepackage{microtype}		
\usepackage{float}			

\usepackage{lipsum}		

\usepackage{titlesec}
\titlespacing\section{0pt}{12pt plus 3pt minus 3pt}{1pt plus 1pt minus 1pt}
\titlespacing\subsection{0pt}{10pt plus 3pt minus 3pt}{1pt plus 1pt minus 1pt}
\titlespacing\subsubsection{0pt}{8pt plus 3pt minus 3pt}{1pt plus 1pt minus 1pt}

\title{Design strategies for efficient, fabrication-feasible extreme-ultraviolet~metalens}

\usepackage{authblk}

\author[1,$\dag$,*]{Shiu Hei Lam}
\author[1,$\dag$]{U Abinash Patro}
\author[1,2,3,4]{Jan Rothhardt}
\author[1,2,5]{Thomas Pertsch}

\affil[1]{Institute of Applied Physics, Abbe Center of Photonics, Friedrich Schiller University Jena, Albert-Einstein-Straße 15, 07745~Jena,~Germany}
\affil[2]{Fraunhofer Institute for Applied Optics and Precision Engineering IOF, Albert-Einstein-Straße 7, 07745 Jena, Germany}
\affil[3]{Helmholtz Institute Jena, Fröbelstieg 3, 07743 Jena, Germany}
\affil[4]{GSI Helmholtzzentrum für Schwerionenforschung, Planckstraße 1, 64291 Darmstadt, Germany}
\affil[5]{Max Plank School of Photonics, Hans-Knoell-Str. 1, 07745 Jena, Germany}
\affil[$\dag$]{The authors contributed equally to this work.}
\affil[*]{shiu.hei.lam@uni-jena.de}

\begin{document}

\twocolumn[ 
  \begin{@twocolumnfalse} 
  
  \maketitle

  \begin{abstract}
      The concept of metasurfaces was recently applied to the extreme ultraviolet (EUV) spectral regime, providing a new opportunity for transmissive focusing elements in a regime where materials are highly lossy. The realization of metalenses in the EUV, however, is challenging due to the optical losses and low refractive index contrast of available materials, as well as the larger-than-wavelength periodicity of metaatom arrays imposed by fabrication limits. In this paper, we propose alternative EUV metalens design strategies, including layout schemes and metaatom mapping rules. We demonstrate that the focusing efficiency can be roughly doubled compared with the simple square-lattice design of an EUV metalens purely by using an alternative semi-analytical design approach without reducing the metasurface's minimum feature size. The proposed strategies are generally applicable to metaoptics design for efficiency improvement when metaatoms are lossy or induce diffraction orders.
  \end{abstract}
  
  \vspace{0.5cm}

    \end{@twocolumnfalse} 
]


  \section{Introduction}

Reducing the length scale is a main direction in modern science and technologies. Guided by Ernst Abbe's equation of the diffraction limit, which states the inverse relation between resolution and wavelength, we resorted to increasingly short-wavelength spectra, such as the extreme ultraviolet (EUV), in order to operate at increasingly small length scales. For example, the wavelength of the light used in lithography for the semiconductor industry has decreased drastically over the last half century, from 436 nm in the visible spectrum to 13.5 nm in the EUV regime \cite{Wu2007extreme}. Similarly, EUV imaging allows for imaging with nanometer-scale resolution with high elemental contrast \cite{Liu2026dose,Chew2026single,Eschen_2025}. 

Despite the advantages of EUV, building the optical path for these applications is challenging due to exotic material properties and the need to operate at nanometer-level precision. For instance, efficient focusing elements, required for both lithography and microscopy, are absent in the EUV. Traditional transmissive bulk lenses would have near-zero energy efficiency due to the low refractive-index contrast between air and the material and the high material optical absorption in the EUV. Currently available transmissive EUV lens elements include Fresnel zone plates and photon sieves \cite{Baez1961,Kipp2001,Cui2024}, which produce higher diffraction orders and reduce the lens efficiency and quality. Therefore, reflective EUV focusing elements constitute the majority of current solutions, but they limit the numerical aperture and heighten the requirement for alignment and aberration correction. To circumvent the problem of an absent transmissive focusing system, the state-of-the-art EUV lithography machines employ a reflective system of 11 mirrors to prepare the focused EUV beam \cite{Wu2007extreme}, while EUV imaging has evolved into a lensless approach, such as coherent diffractive imaging \cite{Eschen2022material}, Fourier transform holography \cite{Zayko2021} and randomized probe imaging \cite{Liu2026dose}.

\begin{figure}[!ht]
\centering\includegraphics[page=1]{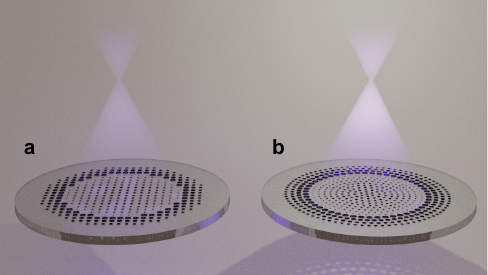}
\caption{Artistic impression of EUV metalenses. EUV metalenses with (a) square-lattice and (b) ring-type layouts are presented. EUV beams, shown in purple, are incident from below the metalenses and are focused by the metalenses.}
\label{fig:overview}
\end{figure}

Recently, the concept of metasurfaces \cite{Lalanne1998, Aieta2012, Khorasaninejad2016} was applied in the EUV regime to realize a transmissive EUV metalens at a 50-nm wavelength, demonstrating sub-micron spot focusing (Fig.~\ref{fig:overview}a) \cite{Ossiander2023}. To adapt to the EUV regime, the proposed EUV metalens differed from traditional metasurfaces in several aspects. While conventional metasurfaces typically consist of particles lying on or being embedded in a substrate, the demonstrated vacuum-guided EUV metalens was a thin silicon membrane patterned with vacuum holes. To accommodate the fabrication limit, the lattice constant of the metaatom array was set to 120 nm, larger than the wavelength. This limit arises because electron beam lithography, a common fabrication tool for metasurfaces with small features, is constrained by the proximity effect and typically yields minimum feature sizes of a few tens of nanometers, while subsequent etching steps also alter the structure's dimensions and profile.

Since the first proposal of an EUV metasurface, there have been theoretical attempts to improve the efficiency of EUV metalenses by engineering the metaatoms, including reducing the feature size to a strictly sub-wavelength region \cite{Zhang2024}, employing complex metaatom shapes \cite{ZarateVillegas2025}, or using stacked holes \cite{Zhao2026}. Although these modifications were proven effective in theory, reducing the minimum feature size or requiring precise control of the hole shape is extremely challenging in practice due to the fabrication-related reasons discussed above. On the other hand, the proposed two-layer approach benefits from a fixed minimum feature size but requires precise alignment of the two layers \cite{Zhao2026}, thereby also increasing the complexity of the fabrication process.

The design task of EUV metalenses is unique because the stringent constraints imposed by the exotic material's optical properties and available fabrication technologies push the solution space beyond the regime conventionally considered suitable for metasurfaces. In such a regime, some basic assumptions or requirements for designing metasurfaces are violated, resulting in a metalens with much lower efficiency than a conventional one at longer wavelengths. Most fundamentally, the larger-than-wavelength lattice constant violates the requirement that the dimension of metaatoms in a metasurface should be sub-wavelength, which is essential to provide an effectively smooth phase modulation of the wave. From the perspective of the Nyquist theorem, the EUV metalenses are undersampled for all numerical apertures (NAs) \cite{Kim2025Nyquist,Kim2025Antialiased}. On the other hand, it is assumed implicitly in the EUV metalenses' design, just as in the standard metalens design method, that the transmittances of metaatoms are uniform, while in practice, the highest metaatom transmittance in EUV metalenses is typically multifold of the lowest one \cite{Ossiander2023,ZarateVillegas2025,Zhao2026}. Furthermore, the fact that the metaatoms library often does not support a $2\pi$ phase at the chosen silicon membrane thickness \cite{Ossiander2023,ZarateVillegas2025}, due to the limited refractive index contrast between silicon and vacuum, is also contrary to the design assumption of a phase-modulating metasurface. 

In this paper, we approach the design of EUV metalenses as an unconventional metasurface design problem, accounting for the aforementioned uncommon material and metaatom properties specific to this spectral regime. We discuss approaches to improve EUV metalens efficiency with the consideration of the fabrication limit of the currently available technologies, where we restrict our metaatoms to be circular holes with a minimum feature size, including hole diameters and gap widths, of 20 nm, which is the same value as in the experimental demonstration by \cite{Ossiander2023}. We discuss alternative design approaches in two directions: the layout scheme and the metaatom mapping rule. The discussion of the layout scheme addresses the violation of the subwavelength sampling assumption and demonstrates how significantly it affects the efficiency of EUV metalenses. We discuss and propose alternative layout schemes for square-lattice EUV metalenses, including the hexagonal-lattice and ring layout schemes (Fig.~\ref{fig:overview}b). The discussion on the metaatom mapping rule addresses violations of the assumptions of uniform metaatom transmittance and the absence of higher diffraction orders. We propose alternative mapping rules to select a metaatom for a target phase at a location on a metasurface, based on preferences for different figures of merit. Although some proposed alternative design approaches were also employed in previous work on conventional metasurface design, these approaches were employed with different motivations, and the effectiveness of the approaches on the performance differs qualitatively due to the unique environment in the EUV aforementioned \cite{Moreno2026,Kim2025Antialiased,Shi2025,Park2024,She2018,Wang2017,Arbabi2020}. Our design strategies are semi-analytical and can therefore be adapted to larger metasurfaces when full numerical simulation becomes impractical. Furthermore, although our study focuses on an EUV metasurface, our proposed strategies are generally applicable to metaoptics design for metaatoms in the diffractive regime and for lossy materials.

\begin{figure}[!ht]
\centering\includegraphics[page=10]{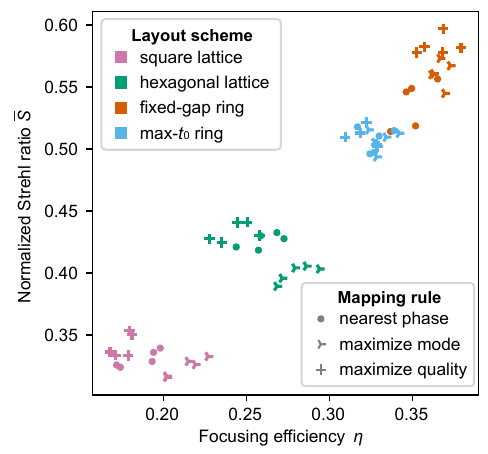}
\caption{A summary of the performance of NA 0.30 EUV metalenses, designed with different layout schemes (square-lattice, hexagonal-lattice, fixed-gap ring and max-$t_0$ ring layouts) and mapping rules ("nearest phase", "maximize mode" and "maximize quality") proposed in this paper. Both the normalized Strehl ratio $\bar{S}$ and the focusing efficiency $\eta$ can differ by nearly a factor of 2 depending on the design method.}
\label{fig:NA0.3}
\end{figure}

In section \ref{sec:layout}, we first discuss the difference between square and hexagonal lattices based on the unit cells and propose two classes of ring-type layout schemes, in which both the hole diameters and the lattice constants of the constituting metaatoms can be varied. We demonstrate, through full-wave simulations, that switching from the square lattice to the hexagonal lattice, and then from the hexagonal lattice to ring-type layouts, significantly improves the performance of the EUV metalens. The differences in the characteristics of various layout schemes are discussed based on their coherent scattering behavior. In section \ref{sec:mapping}, we introduce different mapping rules, "nearest phase", "maximize mode" and "maximize quality", and demonstrate with full-wave simulation that extra performance enhancement can be obtained by selecting a suitable mapping rule. When employed together, the alternative layout schemes and mapping rules can provide nearly a twofold improvement in metalens efficiency (Fig.~\ref{fig:NA0.3}).

\section{Layout scheme}
\label{sec:layout}

We considered an EUV metalens made from a free-standing membrane of crystalline silicon with a refractive index taken from Palik \cite{Palik1998handbook}. The wavelength was set to 46 nm (27 eV), which is around the eleventh harmonic of high harmonic generation when a ytterbium-doped fiber laser source is used in the cascaded frequency conversion scheme \cite{Klas2021,Klas2016,Hilbert2020}. The thickness of the silicon membrane was set to 280 nm, so that a solid membrane introduces a phase shift of slightly more than $2\pi$ for transmitted light relative to free space.

\subsection{Lattice-type layouts}
\label{sec:lattice}

\begin{figure}[!htbp]
\centering\includegraphics[page=2,scale=0.95]{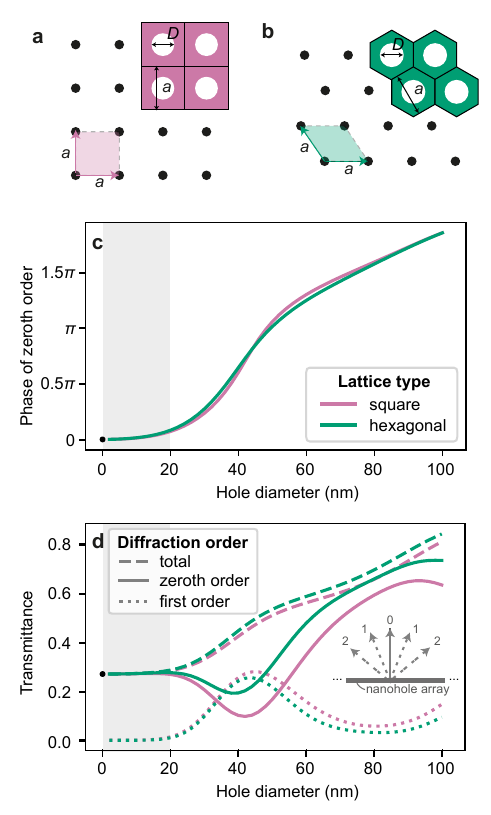}
\caption{The lattice structures and nanohole metaatom mapping for the square- and hexagonal-lattice layout schemes with a fixed lattice constant $a$ of 120 nm. The lattice structures are presented in (a) for the square lattice and (b) for the hexagonal lattice, where the black dots locate the lattice points. The arrows in both subfigures represent the lattice vectors of their respective lattices, each with a length of 120 nm. $2\times 2$ unit cells of nanohole metaatom are presented for each lattice scheme on the right top of (a) and (b), showing the nanoholes with diameter $D$ in white color and the silicon membrane in pink or green. (c) presents the phases of the zeroth diffraction order transmitted light, when light propagates through periodic arrays of nanoholes in the square and hexagonal lattices, as a function of hole diameter. (d) presents the transmittance of the zeroth and first diffraction orders, for both lattice types. The inset of (d) illustrates the presence of multiple diffraction orders in the transmitted light of the nanohole arrays. The total transmissions, the sum of the transmittances of all diffraction orders, are also presented. The colors differentiate the lattice types, as in (c). A data point at zero diameter, corresponding to a unit cell without a hole, is presented in both (c) and (d) and is shared by the square- and hexagonal-lattice metaatom libraries.  The metaatom libraries cover only holes with diameters of 20 to 100 nm and the zero-diameter data point, excluding the gray-shaded region in (c) and (d), to enforce a minimum feature size of 20 nm. The data were obtained with COMSOL. The phases were defined such that the phase of the case without a hole is zero.}
\label{fig:lattice}
\end{figure}

Fig.~\ref{fig:lattice} presents a comparison between nanohole arrays in (a) the square and (b) the hexagonal lattice with a fixed lattice constant $a$ of 120 nm. Data for hole diameters below 20 nm are also presented for completeness, but are excluded from the metalens design study to impose the minimal feature size. The comparison of the transmitted phase of nanohole arrays with different diameters in Fig.~\ref{fig:lattice}c shows that the phase dependence on hole diameter and the overall phase coverage of the metaatom library are similar regardless of the lattice type. However, the total transmission, as presented in Fig.~\ref{fig:lattice}d, is slightly higher for the hexagonal lattice than for the square lattice, whereas the transmittance in the zeroth order is significantly higher for all hole diameters above 20 nm with the hexagonal lattice. The phase-averaged total transmission is 0.50 for square-lattice metaatoms and 0.53 for hexagonal-lattice ones, while the phase-averaged zeroth-order transmittances are 0.30 and 0.40 for square and hexagonal lattices, respectively. The increased total transmission can be attributed to the hexagonal lattice's higher packing factor, which reduces the fraction of lossy silicon in the metasurface. On the other hand, the hexagonal lattice has a larger reciprocal lattice compared to the square lattice of the same lattice constant, which favors the zeroth diffraction order over the higher orders. As shown in Fig.~\ref{fig:lattice}d, the total transmittance of the first orders is higher for the square lattice than for the hexagonal lattice. 

With the conventional design method of phase-modulating metasurface using the local periodic approximation, only the zeroth order contributes to the useful output, whereas the higher orders are considered stray light. The result predicted that simply switching from the square to the hexagonal lattice can increase the energy efficiency of a silicon EUV metalens by one-third without reducing the pitch size. Furthermore, a hexagonal-lattice design should benefit from its higher-order symmetry compared to a square-lattice design for applications of focusing elements, which are expected to be rotationally symmetric. A rigorous evaluation, in which entire metalenses were simulated with an FDTD solver, will be presented later in section \ref{sec:performance}, demonstrating the lens performance improvement brought by a switch from a square lattice to a hexagonal lattice.

\subsection{Ring-type layouts}
\label{sec:ring}

\begin{figure}[!htbp]
\centering\includegraphics[page=3]{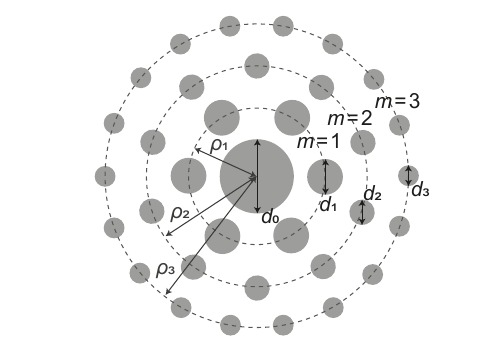}
\caption{Schematic of a ring-type layout. A ring-type layout is defined by the ring radius $\rho_m$, the diameters $d_m$ of holes and the total number of holes $N_m$ of each ring $m$. In the example, the number of holes in each ring, $N_1$, $N_2$, $N_3$, is 6, 10, 18, respectively.}
\label{fig:ring_schmatics}
\end{figure}

\begin{figure}[!htbp]
\centering\includegraphics[page=4]{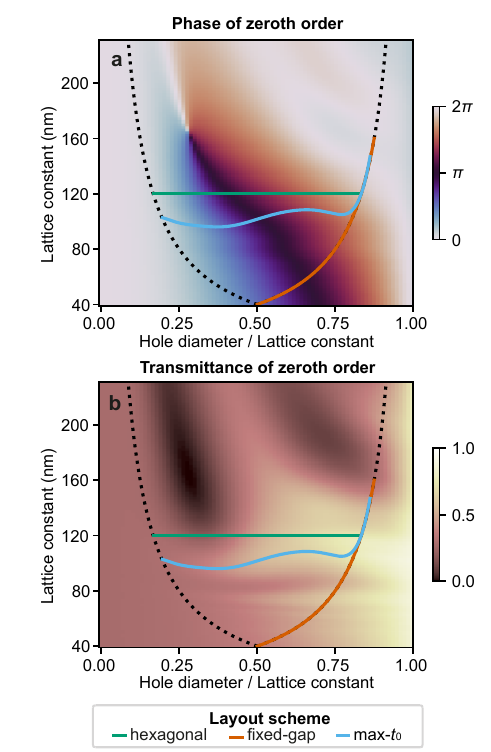}
\caption{Metaatoms libraries for the hexagonal-lattice and ring-type layout schemes in the two-dimensional parameter space. (a) presents the phase of zeroth diffraction order transmitted light, when light propagates through periodic arrays of nanoholes in the hexagonal lattices as a function of the lattice constant and the ratio of hole diameter to lattice constant. (b) presents the transmittance of the zeroth diffraction order, in correspondence to (a). The path that defines the selection of metaatoms for each layout scheme is shown on the map in both (a) and (b), as labeled in the legend at the bottom of the figure. The green path marks the metaatom library with a fixed lattice constant at 120 nm, corresponding to the hexagonal-lattice scheme presented in Figs.~\ref{fig:lattice}c and \ref{fig:lattice}d. The brown and light blue paths correspond to the fixed-gap and the max-$t_0$ schemes of the ring-type layout, respectively. The areas enclosed by the two dotted lines in both (a) and (b) mark the allowed parameter space subject to a minimum feature size of 20 nm. The data were obtained with FDTD Lumerical.}
\label{fig:ring}
\end{figure}

Following the success of the hexagonal-lattice layout, which predicts improved lens performance due to a denser packing of metaatoms compared to the square-lattice layout, it is natural to search for layout schemes that further increase the packing density. Since the hexagonal arrangement has reached the highest packing efficiency for lattice-type arrangements, one must abandon the lattice in order to achieve an even higher packing factor. With the consideration of the azimuthal-angle symmetry that is expected for general lens design, we propose a ring-type layout as illustrated in Fig.~\ref{fig:ring_schmatics}. Adapting a ring-type layout not only increases the packing density of metaatoms but also further improves the rotational symmetry of a metalens as compared to a lattice-type layout.

In a ring-type layout, a ring of metaatoms is referred to by its indices $m = 1,\dots, M$, where $M$ is the total number of rings in a ring-type layout. All metaatoms in the $m$-th ring with ring radius $\rho_m$ have the same hole diameter $d_m$ and lattice constant $a_m$. Given $\rho_0=0$ and $a_0$ being the lattice constant of the center metaatom, the ring design should fulfill the following equations such that the spacing of holes approximates their lattice constant $a_m$ in the unit cells:
\begin{equation}
\label{eq:interring}
\rho_m-\rho_{m-1} = (a_m + a_{m-1})/2, \qquad m = 1, …, M,
\end{equation}
\begin{equation}
\label{eq:intraring}
\frac{N_m}{2} = \lfloor \pi/\sin^{-1}\left(\frac{a_m}{2\rho_m}\right)/2 \rfloor, \qquad m = 1, …, M.
\end{equation}
$N_m$ is the number of holes in the $m$-th ring. The $\lfloor\ \rfloor$ operator rounds down to the nearest integer. Eqs.~(\ref{eq:interring}) and (\ref{eq:intraring}) address the inter-ring and intra-ring holes separation, respectively. Eq.~(\ref{eq:intraring}) rounds $N_m$ to an even number, such that the ring layout has a D2 symmetry, which allows reducing the FDTD simulation domain to a quarter of a whole metalens by setting appropriate boundary conditions and hence speeds up the simulation. $N_m$ is rounded down, instead of rounded off, to ensure the minimum feature size is respected.

The ring layout scheme allows varying the lattice constant in addition to hole diameters, expanding the parameter space from one dimension (Figs.~\ref{fig:lattice}c  and \ref{fig:lattice}d) to two dimensions (Figs.~\ref{fig:ring}a and \ref{fig:ring}b). To perform metaatom mapping using the local periodic approximation, hexagonal unit cells were used instead of square ones, as their higher packing density better represents metaatoms in a highly packed environment. Although one can select any set of metaatoms in the two-dimensional parameter space for the ring layout to cover the phase space, a set that is continuous in the parameter space, i.e., a continuous path, is more robust for fabrication and better aligns with the local periodic assumption. We defined two metaatom selection approaches, namely the fixed-gap approach and the maximum-zeroth-order-transmittance (max-$t_0$) approach, whose metaatom selection paths are shown in Fig.~\ref{fig:ring} by brown and blue curves, respectively. The fixed-gap scheme minimizes the fraction of silicon in the metasurface by fixing the gap distances of holes to 20 nm, the minimum feature size. The max-$t_0$ scheme selects the metaatoms that cover approximately $2\pi$ phase with the highest zeroth-order transmittance along a continuous path within the region allowed by the minimum feature size. The minimum feature size of the ring layout was ensured by restricting the metaatom selection path to the parameter space in which the minimum hole and gap sizes are 20 nm. Alternative visualizations of the zeroth diffraction order transmission coefficients of each layout scheme in the complex plane are provided in Fig.~\ref{fig:metaatom_t_coeff}, which aids the comparison of the phase coverage and transmittance between different layout schemes.

\subsection{Metalens performance of different layout schemes}
\label{sec:performance}
To rigorously compare the performance of metalenses designed with different layout schemes, a series of simulations of entire metalenses was performed. The output electric field profile of each metalens design, which was illuminated by a Gaussian beam with a waist of 2 $\mu$m, was obtained from FDTD simulation using Lumerical, and was propagated to free space with the band-limited angular spectrum method \cite{Matsushima2009}. The lens performance was evaluated in terms of two figures of merit, the normalized Strehl ratio $\bar{S}$ and focusing efficiency $\eta$, where the former is defined as follows:
\begin{equation}
\label{eq:Strehl_ratio}
\bar{S} =
\frac{|\textbf{E}(0,0;z_f)|^2}{|\textbf{E}_\text{ideal}(0,0;z_{f,\text{ideal}})|^2}/
\frac{P(z_\text{out}; \Omega_\text{sim})}{P_\text{ideal}(z_\text{out}; \Omega_\text{sim})}.
\end{equation}
$\textbf{E}(x,y;z)$ is the electric field obtained from the simulation of a lens with light propagating to the $+z$ direction. $z=z_f$ is the plane with the maximum electric field intensity at the optical axis, $(x,y)=(0,0)$. The subscript "ideal" denotes the corresponding data obtained from a simulation of an ideal lossless parabolic lens phase profile. $\bar{S}$ is normalized by the ratio between the output power of the evaluated lens and the ideal lens, such that the energy efficiency of the evaluated lens does not contribute to $\bar{S}$. To be precise,
\begin{equation}
\label{eq:P}
P(z; \Omega) = \frac{1}{2Z_0}{\iint_\Omega |\textbf{E}(x,y;z)|^2 dx\ dy}
\end{equation}
is the power through the area defined by $\Omega$ at plane $z$. $Z_0$ is the wave impedance in free space. In Eq.~(\ref{eq:Strehl_ratio}), $z_\text{out}$ is the output plane, immediately after the lens, whereas $\Omega_\text{sim}$ is the transverse region of the whole simulation domain. 

The second figure of merit, the focusing efficiency, is defined as
\begin{equation}
\label{eq:focusing_eff}
\eta =\frac{P(z_{\eta}; r\leq r_\text{airy})}{P(z_\text{in}; \Omega_\text{sim})}.
\end{equation}
$z_\text{in}$ is the input plane, immediately before the lens. $z_{\eta}$ is the plane with the highest $\eta$, and does not necessarily equal $z_f$ when a lens is aberrated. The denominator integrates the electric field intensity over the whole input plane in the simulation $\Omega_\text{sim}$, whereas the numerator integrates the electric field intensity over the area within the first airy disk minimum, $r_\text{airy} = 0.61\lambda/\text{NA}$. 

The two figures of merit, $\bar{S}$ and $\eta$, were designed to describe the image quality and the energy efficiency of a lens separately. We note that the absolute values of the figures of merit stated in this work might not be directly comparable with other works due to the variation of their definitions.

\begin{figure*}[!htbp]
\centering\includegraphics[scale=0.9,page=5]{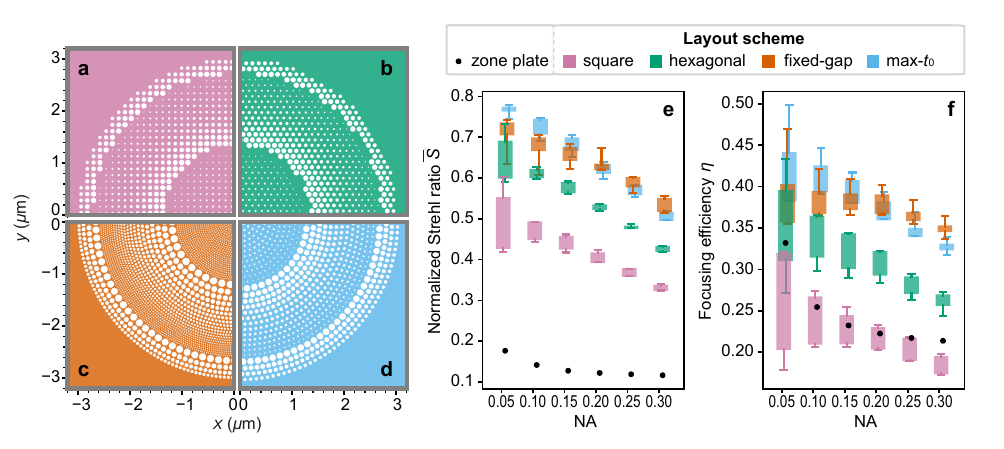}
\caption{Performance comparison of different layout schemes. (a--d) Metalens layouts of NA 0.05 are presented for a global phase shift of 0.6$\pi$ using the four layout schemes: (a) square lattice, (b) hexagonal lattice, (c) fixed-gap ring and (d) max-$t_0$ ring. In all layouts, white represents holes, whereas the colored regions (pink/green/brown/light blue) represent silicon. A quarter of each discussed metalens is sufficient to specify its design, since all discussed metalenses are D2-symmetric. The centers of the metalenses are at $(x,y)=(0,0)$. The right panel summarizes the performance of metalenses obtained from different layout approaches (square lattice, hexagonal lattice, fixed-gap ring, max-$t_0$ ring) as well as the binary zone plate. The (e) normalized Strehl ratio $\bar{S}$ and (f) focusing efficiency $\eta$ of metalenses for different NAs are presented. For each combination of NA and layout approach, five realizations of the metalens were evaluated by varying the global phase $\Phi_g$. The performances of the five realizations in each combination are represented as a box and its whiskers. The caps of the whiskers show the minimum and maximum values among the five realizations, and the lower and upper bounds of the box show the first quartile and third quartile calculated from the values of the five realizations, respectively. The box and whiskers of each optical element are slightly shifted horizontally for readability. The metalenses presented in this figure were designed with the "nearest phase" mapping rule.}
\label{fig:layout}
\end{figure*}

We studied the performance of the metalens with a fixed lens radius $R=3$ $\mu$m at NAs of 0.05, 0.1, 0.15, 0.2, 0.25 and 0.3 (Fig.~\ref{fig:layout}). The phase profile $\phi_\text{lens}$ of a metalens with focal length $f = R/\tan(\sin^{-1}(\text{NA}))$ is given by
\begin{equation}
\phi_\text{lens}(x,y) = -\frac{2\pi}{\lambda} \left(\sqrt{x^2+y^2+f^2}-f  \right) + \Phi_g,
\label{eq:phaseprofile}
\end{equation}
where $\Phi_g$ is a constant global phase of the whole metalens. The phase profiles of NA 0.05 metalenses with different $\Phi_g$ are presented in Fig.~\ref{fig:lens_phase_profile} as an example.

Metalens layouts were obtained by selecting and placing metaatoms to achieve the target phase $\phi_\text{lens}$ in the lens profile. Figs.~\ref{fig:layout}a--\ref{fig:layout}d demonstrate the metalens layouts of the same phase profile (NA = 0.05, $\Phi_g=0.6\pi$) using the different layout schemes proposed in Sections \ref{sec:lattice} and \ref{sec:ring}. The square-lattice (Fig.~\ref{fig:layout}a) and hexagonal-lattice (Fig.~\ref{fig:layout}b) layouts are very similar in terms of the spatial distribution of holes of different diameters because of their similar phase dependence on the hole diameter, as discussed with Fig.~\ref{fig:lattice}b. On the other hand, the packing of holes is denser for the ring-type layouts (Figs.~\ref{fig:lattice}c and \ref{fig:lattice}d), where the fixed-gap layout shows the highest packing of holes.

For an ideal phase mask, the global phase $\Phi_g$ in Eq.~(\ref{eq:phaseprofile}) can be selected arbitrarily without changing the lens performance. However, for a metasurface, the global phase affects the placement of metaatoms, which can alter the lens performance, especially when the metaatoms are far from being ideal phase modulators. To provide a comprehensive understanding and a fair comparison of the layout schemes, five metalens layouts were obtained by varying the global phase $\Phi_g$ for each combination of layout scheme and numerical aperture, yielding a total of 120 metalens designs. As an example, the layouts of NA 0.05 metalenses designed with the square-lattice scheme with different $\Phi_g$ are provided in Fig.~\ref{fig:square_layouts}. 

The normalized Strehl ratio $\bar{S}$ and the focusing efficiency $\eta$, as described in Eq.~(\ref{eq:Strehl_ratio}) and (\ref{eq:focusing_eff}), were evaluated and plotted in Figs.~\ref{fig:layout}e and \ref{fig:layout}f, respectively. The performances of ideal binary zone plates are also plotted to provide a baseline for the current transmissive EUV lens. (Details on the modeling of the zone plates are provided in the supplementary information, Section \ref{sec:zone_plate}.) The data are reduced to box plots to highlight the performance characteristics of different layout schemes. The full data are provided in Fig.~\ref{fig:layout_performance_detailed}, which shows the figures of merit, $\bar{S}$ and $\eta$, for each of the 120 simulated metalens.

The lens performances of all layout schemes, as well as the zone plates, worsen as the NA increases, due to a steeper phase gradient in the lens phase profile and, consequently, more severe undersampling. Both $\bar{S}$ and $\eta$ increase substantially when the layout scheme evolves from the square-lattice scheme to the hexagonal-lattice scheme, and even more when changing to the ring-type layout schemes, including the fixed-gap and max-$t_0$ schemes, for all investigated NAs. Considering the maximum performance of each set, $\bar{S}$ is increased by 22\%--29\% and $\eta$ by 34\%--38\% for different NAs when switching from the square-lattice scheme to the hexagonal-lattice scheme. When switching from a hexagonal-lattice layout to a ring-type layout, the maximum $\bar{S}$ and $\eta$ are improved by a further 6\%--29\% and 15\%--34\%, respectively, for different NAs. Switching from the square-lattice layout to the ring-type layout yields an overall improvement of 29\%--64\% in the highest $\bar{S}$ and 56\%--84\% in the highest $\eta$ for different NAs. The result suggests that our alternative layout schemes, without increasing fabrication requirements, can provide EUV metalens performance enhancements comparable to or exceeding those of previous work \cite{Zhao2026,ZarateVillegas2025}.

Reflected by the length of the boxes and whiskers in Figs.~\ref{fig:layout}e and \ref{fig:layout}f, both $\bar{S}$ and $\eta$ vary depending on the global phase $\Phi_g$ as predicted. The variation is largest at small NA and generally decreases gradually with increasing NA. The dependence on the NA can be explained by a biased population of phase values across the lens profile, which translates into a biased population of hole diameters on the metalens. Gaussian illumination magnifies the bias. For example, the lattice-type metalens with global phase $\Phi_g = 1.0\pi$, corresponding to layouts with a large population of low-transmittance holes with a diameter around 40 to 50 nm at the metalens center, have significantly lower focusing efficiency at low NA than their counterparts with high-transmittance holes at the center ($\Phi_g = 0.2\pi$ or $1.8\pi$). In fact, the metalenses with $\Phi_g = 0.2\pi$ or $1.8\pi$ generally show the highest focusing efficiency, but the advantage narrows with increasing NA (Fig.~\ref{fig:layout_performance_detailed}b). For ring-type layouts, the dependence of the performance on the global phase is slightly less prominent since the ring-type layouts have more uniform transmittance for metaatoms in their libraries (Fig.~\ref{fig:metaatom_t_coeff}).

Despite the large deviation of the ring layout from the lattice arrangement, the high performance of the ring-type layouts suggested that the metasurface mapping method with the local periodic approximation remains valid for ring-type EUV metalens of propagation phase. It is likely that the high absorption of crystalline silicon makes a contribution to relaxing the requirement for the local periodic approximation method to be applicable by hindering the inter-metaatom interactions (Fig.~\ref{fig:derivative_t_a}).

\begin{figure}[!htbp]
\centering\includegraphics[page=7]{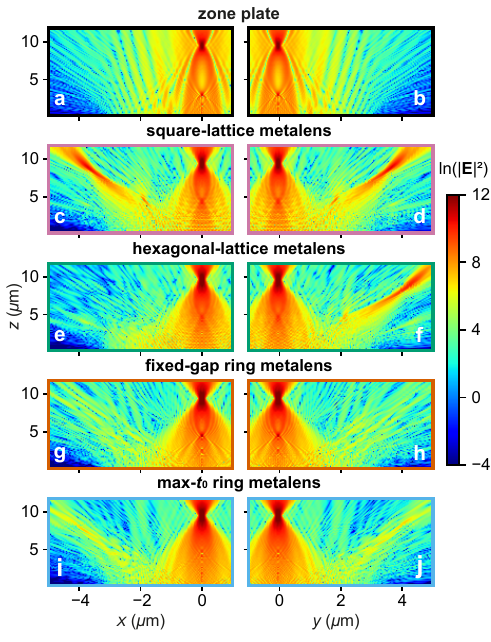}
\caption{Propagation of light after (a, b) binary zone plate and metalenses designed with different layout schemes: (c, d) square lattice, (e, f) hexagonal lattice, (g, h) fixed-gap ring, and (i, j) max-$t_0$ ring. The electric field intensity is visualized in logarithmic scale for $x-z$ (left) and $y-z$ (right) planes crossing the optical axis ($z$-axis). The centers of the output surfaces of the binary zone plate and the metalenses are at $(x,y,z)=(0,0,0)$. The NA of all lenses is fixed at 0.30, while the global phase $\Phi_g$ of the metalenses is fixed at $1.8\pi$.}
\label{fig:prop}
\end{figure}

To observe the different natures of the metalenses designed with different layout schemes, the propagation of light after the metalens is visualized for each layout scheme, for NA 0.30 and $\Phi_g=1.8\pi$ in Fig.~\ref{fig:prop}. The propagation of light after a binary zone plate of the same NA is also provided for reference in Figs.~\ref{fig:prop}a and \ref{fig:prop}b. The visualization of the electric field intensity in the corresponding focal planes can be found in Fig.~\ref{fig:focalplane} of the supplementary information. 

The zone plate and the four presented metalenses all focus light along the optical axis, around their focal length at $z=9.54$ $\mu$m. The characteristic first-diffraction order foci are clearly visible for the square-lattice metalens in the $x-z$ (Fig.~\ref{fig:prop}c) and $y-z$ planes (Fig.~\ref{fig:prop}d) and the hexagonal-lattice metalens in the $y-z$ plane (Fig.~\ref{fig:prop}f). A first diffraction order is absent in the $x-z$ plane of the hexagonal lattice (Fig.~\ref{fig:prop}e) because its reciprocal lattice vector does not cross the $x-z$ plane (Fig.~\ref{fig:focalplane}c). The first diffraction order occurs at a slightly higher polar angle for the hexagonal-lattice metalens than for the square-lattice metalens due to the smaller reciprocal lattice of the hexagonal lattice. For the ring-type-layout metalenses, the fixed-gap metalens does not show any clear secondary focus at higher polar angles, while weak secondary foci at a similar polar angle as the first-diffraction order foci in the hexagonal-lattice metalens are observed for the max-$t_0$ ring metalens. It should be noted that due to the layout symmetry, the "foci" of the max-$t_0$ ring metalens when observed in the $x-y$ plane are rings, instead of discrete focus spots as in the square- or hexagonal-lattice metalenses (Fig.~\ref{fig:focalplane}). Furthermore, zone-plate-type higher-order foci of varying strength can be observed along the optical axis at approximately half the focal length for all four metalenses, with the one in the fixed-gap metalens being the most prominent.

While it is well-understood that the lattice-type metalenses with a lattice constant larger than the wavelength would have extra foci at their higher diffraction orders, the behavior of the ring-type layout metalenses is less straightforward to understand. In the attempt to describe the interference of light due to the ring-type layouts from a similar perspective as the grating effect, we estimate the coherent scattering behavior of a layout type by introducing the quantity $L(k_r)$, which we refer to as the layout-induced coherence factor:
\begin{equation}
L(k_r) = \left\langle \int_0^{2\pi} \frac{1}{N}|\textbf{E}_s^0(\textbf{k})|^2\left|\sum_{j=1}^N \exp(-i\textbf{k}_\parallel\cdot\textbf{r}_j) \right|^2\ d\alpha \right\rangle.
\label{eq:Lk}
\end{equation}
$N$ is the total number of metaatoms in a layout. $\textbf{r}_j=(x_j,y_j)$ is the position vector of the $j$-th metaatom from the center of the metalens. $\textbf{k}_\parallel$, in polar coordinates $(k_r, \alpha)$, is the transverse wavevector which defines the scattering polar angle $\theta = \sin^{-1}(k_r/k_0)$. The exponent in Eq.~(\ref{eq:Lk}) describes the coherence of scattering due to the placement of the metaatoms for a layout scheme of interest. The factor of $1/N$ scales the input power per particle with the number of particles, such that the total input power is conserved. $|\textbf{E}_s^0(\textbf{k})|^2$ is the scattering intensity of a single metaatom. Denoted by the angle brackets $\left<\ \right>$, the integral in Eq.~(\ref{eq:Lk}) is averaged over the layouts of the same category. 

Although it resembles the specular contribution to the average scattered intensity of a disorder metasurface \cite{Lalanne2025}, the integrand in Eq.~(\ref{eq:Lk}) is not a quantitative description of the specular scattering of the metalens in our analysis, since inhomogeneous metaatoms are involved. $|\textbf{E}_s^0(\textbf{k})|^2$, estimated here by the average far field of metaatoms for simplicity, serves only as a window function to filter out irrelevant angular spectrum. Description of the calculation of the far field of the holes can be found in the supplementary information Section \ref{sec:ff} and Figs.~\ref{fig:ff_2d} and \ref{fig:ff_1d}. $L(k_r)$ shall allow us to predict the occurrence of secondary foci due to coherent scattering at a particular polar angle $\theta$, but its intensity should not be interpreted as the strength of a predicted secondary focus.

\begin{figure}[!htbp]
\centering\includegraphics[page=6]{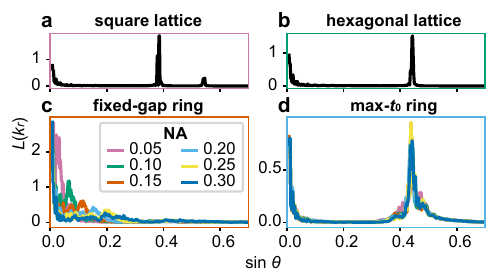}
\caption{Prediction of coherent scattering in the polar angle $\theta$ with the layout-induced coherence factor $L(k_r)$, as defined in Eq.~(\ref{eq:Lk}). (a--d) presents $L$ for square-lattice, hexagonal-lattice, fixed-gap ring and max-$t_0$ ring layout schemes, respectively. For lattice-type layouts, $L$ is independent of the NA. For ring-type layouts, $L(k_r)$ is evaluated for each investigated NA.}
\label{fig:Lk}
\end{figure}

Fig.~\ref{fig:Lk} presents the layout-induced coherence factor $L(k_r)$ for each proposed layout scheme and each NA, evaluated numerically. Each $L(k_r)$ was averaged over the five realizations of the phase profile by shifting $\Phi_g$. As one would expect for lattice arrangements, peaks can be found at $k_r = 0.38$ and 0.54 for square-lattice layouts (Fig.~\ref{fig:Lk}a), and at $k_r = 0.44$ for hexagonal-lattice layouts (Fig.~\ref{fig:Lk}b), which corresponds to the first and second diffraction orders of the sqaure lattice and first diffraction order of the hexagonal lattice, respectively. For the max-$t_0$ ring layouts (Fig.~\ref{fig:Lk}d), a peak was found at $k_r=0.44$. The origin of the peak can be traced to the construction of the metaatom library for the max-$t_0$ ring scheme, in which the lattice constants are populated within a narrow range of 96 to 147 nm. As already seen in Figs.~\ref{fig:prop}i and \ref{fig:prop}j, light is focused at the corresponding polar angle by the max-$t_0$ ring metalens.

Unlike the max-$t_0$ scheme, the fixed-gap scheme (Fig.~\ref{fig:Lk}c) gives nearly no signal for $k_r>0.3$, which aligns with the absence of foci at large polar angles for the fixed-gap metalenses observed in Figs.~\ref{fig:prop}g and \ref{fig:prop}h. Intriguingly, $L(k_r)$ for fixed-gap metalenses exhibits a prominent peak at small $k_r$, a feature not observed in other layout schemes. The peak shifts to higher $k_r$ and weakens as NA increases. This unique feature is a direct consequence of the fixed gaps between holes: since the phases of metaatoms increase with hole diameter while the gap is fixed, the center-to-center distance of the metaatoms in the ring layout shortens or lengthens as the phase decreases or increases, respectively. Consequently, the fixed-gap ring layout has metaatom spacings correlated with the lens's phase profile. This arrangement highly resembles the construction of a zone plate, in which transparent rings are situated where the lens profile has the same phase. The zone-plate-like behavior of the fixed-gap metalenses is distinguished, as already illustrated by the strong higher-order foci along the optical axis in Figs.~\ref{fig:prop}g and \ref{fig:prop}h. It is revealed that a fixed-gap metalens has the hybrid nature of a conventional metalens and a zone plate. The weakening of $L(k_r)$ with increasing NA suggests that the zone-plate-like feature of the fixed-gap metalens is reduced with increasing NA, which provides a possible explanation for why the performance of the fixed-gap metalenses overtake the max-$t_0$ ring metalenses as NA increases.

\section{Metaatom mapping rules}
\label{sec:mapping}
\begin{figure}[!htbp]
\centering\includegraphics[page=8]{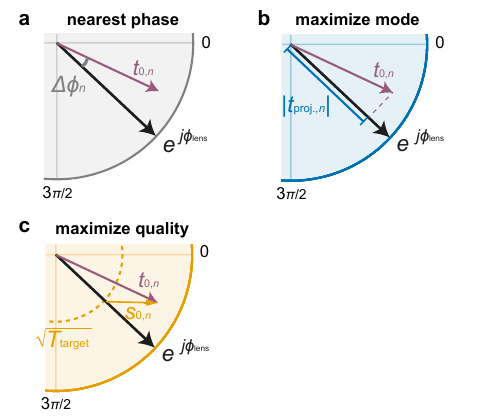}
\caption{Visualization of the quantities involved in the three mapping rules, (a) "nearest phase", (b) "maximize mode and (c) "maximize quality", on the complex planes of transmission coefficients of metaatoms.}
\label{fig:mapping_concept}
\end{figure}

From the strong dependence of the metalens performance on the global phase $\Phi_g$, it became evident that the design of EUV metasurfaces with a non-ideal metaatom library, i.e., a library consisting of metaatoms with higher diffraction orders or varying magnitudes of zeroth-diffraction-order transmittances or a library that does not have $2\pi$-phase coverage, requires special consideration. In addition to varying the global phase $\Phi_g$, the mapping can also be altered by changing the mapping rules. Previously, we have employed a mapping rule ("nearest phase"), which selects the metaatom $n$ for a position $(x,y)$ in a metalens by minimizing the phase error $\Delta\phi_n$ (Fig.~\ref{fig:mapping_concept}a), meaning,
\begin{equation}
	n^*\ \text{minimizes}\ 
	\left|\Delta\phi_n\right|
	\label{eq:nearest_phase},
\end{equation}
where
\begin{equation}
\Delta\phi_n=\arg\{t_{0,n}\cdot \exp(-j[\phi_\text{lens}(x,y)])\}.
\end{equation}
$t_{0,n}$ is the complex transmission coefficient of the zeroth diffraction order of the metaatom $n$. 

Nonetheless, when one would like to maximize the conversion to the target complex field, one can define an alternative mapping rule ("maximize mode") based on mode-matching (Fig.~\ref{fig:mapping_concept}b), which means to
\begin{equation}
n^*\ \text{maximizes}\ 
\left|t_{\text{proj.},n}\right|,
\label{eq:max_mode}
\end{equation}
where
\begin{equation}
|t_{\text{proj.},n}|
=\text{Re}(t_{0,n}\cdot \exp(-j[\phi_\text{lens}(x,y)])).
\end{equation}

Otherwise, when one aims at maximizing the quality of the output light field regardless of the energy efficiency, which means to minimize the deviation from the target phase profile of uniform transmittance, one can define a mapping rule ("maximize quality") as (Fig.~\ref{fig:mapping_concept}c):
\begin{equation}
n^*\ \text{minimizes}\ 
\left[|s_{0,n}|^2 + S_n\right],
\label{eq:max_quality}
\end{equation}
where
\begin{align}
s_{0,n}
=&\ t_{0,n} - \sqrt{T_\text{target}}\ \exp(-j[\phi_\text{lens}(x,y)]),\\
S_n
=&\ T_{\text{total},n} - |t_{0,n}|^2.
\end{align}
$T_\text{target}$ is defined for the entire metalens and corresponds to the target mode-matching efficiencies. The optimal $T_\text{target}$ that minimizes Eq.~(\ref{eq:max_quality}) is to be searched by sweeping $T_\text{target}$. $|s_{0,n}|^2$ corresponds to the deviation from the target field of uniform transmittance contributed by the zeroth-order transmittance. $S_n$ are the contributions of higher diffraction orders.

The three definitions of mapping rules, "nearest phase" [Eq.~(\ref{eq:nearest_phase})], "maximize mode" [Eq.~(\ref{eq:max_mode})] and "maximize quality" [Eq.~(\ref{eq:max_quality})], are equivalent when the metaatom library is ideal, meaning that all metaatoms have uniform transmittance, the lattice constant is subwavelength, and the library has $2\pi$ phase coverage. As the metaatoms deviate further from being ideal, a greater difference between the metalenses obtained from the three rules is expected.

\begin{figure*}[!htbp]
\centering\includegraphics[scale=0.9,page=9]{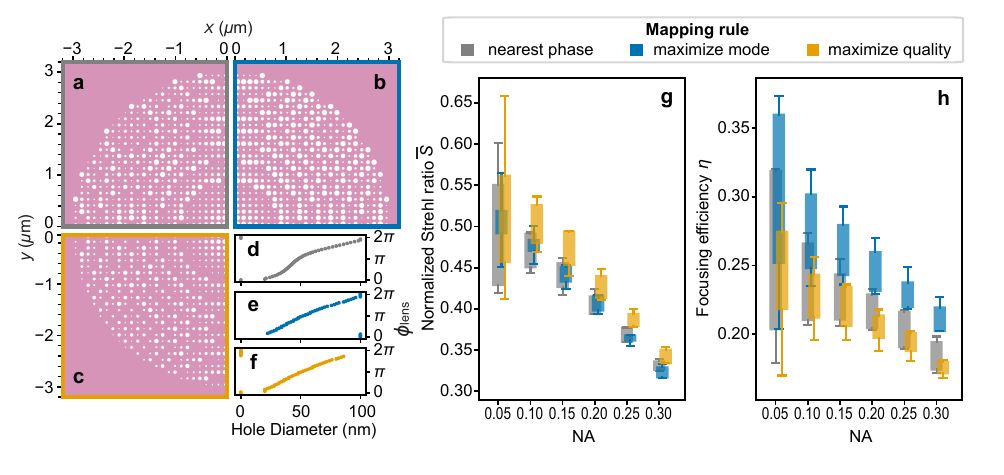}
\caption{Performance comparison for different metaatom mapping rules with square-lattice layout. Metalenses with NA = 0.3 and $\Phi_g = 0.2\pi$ were obtained with mapping rules "nearest phase", "maximize mode", and "maximize quality", with their layouts presented in (a--c), respectively. (d--f) presents the mapping from phase to hole diameter when designing metalenses with each mapping rule, respectively. The right panel summarizes the performance of different mapping rules for square-lattice layout and varying NAs, where (g) shows the normalized Strehl ratio $\bar{S}$ and (h) the focusing efficiency $\eta$. The definitions of the boxes and whiskers follow Fig.~\ref{fig:layout}.}
\label{fig:mapping}
\end{figure*}

Figs.~\ref{fig:mapping}a--\ref{fig:mapping}c present the metalens layout designed with the square-lattice layout scheme, with NA 0.30 and $\Phi_g = 0.2\pi$, mapped using the three different rules. The three layouts closely resemble each other. Figs.~\ref{fig:mapping}d--\ref{fig:mapping}f shows the hole diameters that are to be selected for different phases according to the three different rules, respectively. The most pronounced difference in the mappings lies at the phase gaps, i.e., around phases 0 and $2\pi$. The "maximize mode" rule prefers the large holes, which have the highest transmittance, while the "maximize quality" rule prefers unit cells without holes, which minimize higher diffraction orders. The "nearest phase" rule lies between the other two rules.

Furthermore, with the "nearest phase" rule, it is sampled more densely around hole diameters of 40 to 60 nm due to greater phase variation in this region. On the contrary, both the "maximize mode" rule and the "maximize quality" rule tend to avoid the region of diameters from 40 to 60 nm because of the low transmittance of the zeroth diffraction order and strong contributions from higher diffraction orders. The avoidance shows as a more linear sampling in Figs.~\ref{fig:mapping}e and \ref{fig:mapping}f compared to Fig.~\ref{fig:mapping}d.

Figs.~\ref{fig:mapping}g and \ref{fig:mapping}h compare the normalized Strehl ratio $\bar{S}$ and the focusing efficiency $\eta$ of square-lattice metalenses obtained with different mapping rules, evaluated for different NAs. Similar to the metalens obtained with the "nearest phase" rule, the performance of metalenses obtained with the "maximize mode" rule and the "maximize quality" rule worsens with increasing NA and depends on the global phase $\Phi_g$, with the $\Phi_g$-dependence reducing with increasing NA. The "maximize quality" rule always provides the metalens with the highest $\bar{S}$, improving the highest performance by 4\%--9\% for different NAs compared to the "nearest phase" rule. On the other hand, the "maximize mode" rule always provides the metalens with the highest $\eta$, improving the highest focusing efficiency by 14\%--17\% for different NAs compared to the "nearest phase" rule.

The same analysis as in Fig.~\ref{fig:mapping} for the square-lattice metalenses was also performed for metalenses designed with hexagonal-lattice, fixed-gap ring and max-$t_0$ ring layout schemes (Figs.~\ref{fig:mapping_hex}--\ref{fig:mapping_fp}). The "maximize mode" rule yields metalenses with the highest $\eta$, while the "maximize quality" rule yields metalenses with the highest $\bar{S}$ in general. The difference in the lens performance between different mapping rules is smaller for the ring-type layouts than for the lattice-type layouts because the metaatom libraries of the ring-type schemes are closer to an ideal library. More discussion is provided in the supplementary information, Section \ref{sec:mappingrules}.

It should be noted that the improvement in one figure of merit by employing a mapping rule is complemented by a reduction of the other figure of merit. The inverse relation is most pronounced for the lattice-type metalens, as shown in Figs.~\ref{fig:NA0.3} and \ref{fig:summary_allNA}, which summarize for each NA, the performance of metalenses designed with all combinations of the discussed layout schemes, mapping rules and global phase shifts. The mapping rule should be selected based on the intended application of the metalens. To achieve a customized balance between the phase quality and the energy efficiency, a tailored mapping rule can be defined as a combination, such as a weighted sum, of the "maximize mode" and "maximize quality" rules.

Although the improvement from the mapping rule is small compared to that from a change in layout scheme described in the previous section, the alternative mapping rules can be employed regardless of the layout scheme, and are generally suitable for a phase-modulating metasurface with a non-ideal metaatom library. The improvement brought by the alternative mapping rule is additive to that brought by the layout scheme, meaning a combined implementation of both can yield the best-performing EUV metalens. Compared to the square-lattice metalenses mapped with the "nearest phase" rule, the highest $\bar{S}$ and $\eta$ of an EUV metalens can be improved by 42\%--76\% and 58\%--91\%, respectively, for the NA range of 0.05--0.30 when both an alternative layout scheme and mapping rule are applied.

\section{Conclusion}
In this study, we provide strategies for designing EUV metalens, which differ significantly from conventional metalens design approaches due to substantial material absorption in the EUV spectrum and fabrication-resolution limitations. We proposed the hexagonal-lattice, fixed-gap ring and max-$t_0$ ring schemes as alternative layout schemes to the square-lattice scheme for designing EUV metalenses. By evaluating and comparing the performance of metalenses designed with different layout schemes, we demonstrated the superior energy efficiency and phase quality of the proposed alternative layout schemes. The hexagonal layout scheme can be implemented with minimal effort, but still results in a considerably improved normalized Strehl ratio $\bar{S}$ of 0.73--0.43 and a focusing efficiency $\eta$ of 43\%--27\% for the NA range of 0.05--0.30, corresponding to average improvements of 27\% and 36\%, respectively. The ring-type layouts require a more complex layout algorithm but provide the best lens performance, improving $\bar{S}$ to 0.86--0.60 and $\eta$ to 49\%--38\%, corresponding to average accumulated improvements of 51\% and 86\%, respectively. The characteristics of metalenses with different layout schemes were elucidated by investigating the field propagation and coherence-scattering properties of each scheme. The proposed layout schemes not only improve lens performance but also promote the lens's rotational symmetry.

Furthermore, we proposed alternative metaatom mapping rules, "maximize mode" and "maximize quality", which can be employed to provide an additional improvement in performance on the order of 10\% based on the target application. Considering both the implementation of an alternative layout scheme and mapping rule, $\bar{S}$ and $\eta$ can be improved by 42\%--76\% and 58\%--91\%, respectively, for the NA range of 0.05--0.30.

Although our demonstration was with metalenses of a 3-$\mu$m lens radius, the proposed strategies are semi-analytical and can be scaled to larger lens radii, where rigorous numerical simulations are infeasible because both simulation time and memory scale with metalens size. While the minimum feature size is fixed at 20 nm in this study, the proposed strategies can be employed when only a larger minimum feature size is achievable with the available fabrication tool to improve the efficiency of a metalens design under fixed fabrication constraints. The ring-type layouts are not limited to EUV metalens but also apply to metasurfaces with rotationally symmetric phase profiles, such as corrective optical elements. Furthermore, the hexagonal-lattice layout scheme and the alternative metaatom mapping rule can be employed in general for phase-modulating metasurfaces, such as beam-shapers or computer-generated holograms, in and beyond the EUV regime. Our study not only offers feasible solutions to improve the efficiency of EUV metalenses using existing fabrication technologies, but also provides general insights into the design of metalenses under exotic conditions, where material absorption is severe and the minimum feature size is on the order of the wavelength. 
  
\vspace{1.7em}
\hrule
\vspace{0.7em}
{\bfseries Funding.}\quad This work was partially funded by the Deutsche Forschungsgemeinschaft (DFG, German Research Foundation) through the International Research Training Group (IRTG) 2675 "Meta-Active", project number 437527638 and Collaborative Research Centre (SF13) 1375 "NOA", project number 398816777. This work was also supported by the Free State of Thuringia and the European Social Fund Plus (2025FGR0048). Una Abinash Patro was funded by the German Academic Exchange Service (DAAD) through the Graduate School Scholarship Program. 

\vspace{0.5em}
{\bfseries Disclosures.}\quad The authors declare no conflicts of interest.

\vspace{0.5em}
{\bfseries Supplemental document.}\quad Supplement material for supporting content.
\vspace{1.7em}
\hrule
\vspace{0.7em}

  \bibliography{reference}

\clearpage
\onecolumn

\thispagestyle{empty}
\setcounter{page}{0}
\renewcommand{\thepage}{S\arabic{page}}
\setcounter{figure}{0}
\renewcommand{\thefigure}{S\arabic{figure}}
\setcounter{section}{0}
\renewcommand{\thesection}{S\arabic{section}}

\phantomsection
\addcontentsline{toc}{section}{Supplementary Information}

\begin{center}
  {\LARGE\bfseries Supplementary Information\par}
  {\large\bfseries Design strategies for efficient, fabrication-feasible extreme-ultraviolet~metalens}
\end{center}

\vspace{1cm}

\section{Modeling of the binary zone plates}
\label{sec:zone_plate}
The binary zone plates were modeled as free-standing binary amplitude masks in this work. The radii $r_m$ of the border between alternating 1-and 0-amplitude rings are given by:
\begin{equation}
    r_m^2 = mf\lambda + m^2\lambda^2/4,
\end{equation}
where $f$ is the target focal length, $\lambda$ is the wavelength, and $m$ is the zone number\footnote{Young, M. (1972). Zone plates and their aberrations. \textit{Journal of the Optical Society of America}, \textit{62}(8), 972-976.}. The field propagation after a zone plate was calculated with the same method (the band-limited angular spectrum method) as the metalenses in this work.

\section{Calculation of the far fields of nanoholes}
\label{sec:ff}
The far fields of nanoholes were obtained using Lumerical FDTD. In a simulation, a 5-$\mu$m-thick silicon membrane with a nanohole was illuminated by a total-field scattered-field source. The scattered field of the nanohole was recorded with a 2D field monitor with the same lateral size as the silicon membrane. The far field of the nanohole was then obtained by a near-field-to-far-field transformation of the recorded scattered field.

\section{Performance comparison of different mapping rules for metalenses with hexagonal-lattice, fixed-gap ring and max-$t_0$ ring layout schemes}
\label{sec:mappingrules}
\begin{figure}[htbp]
\centering
\includegraphics[scale=0.8,page=8]{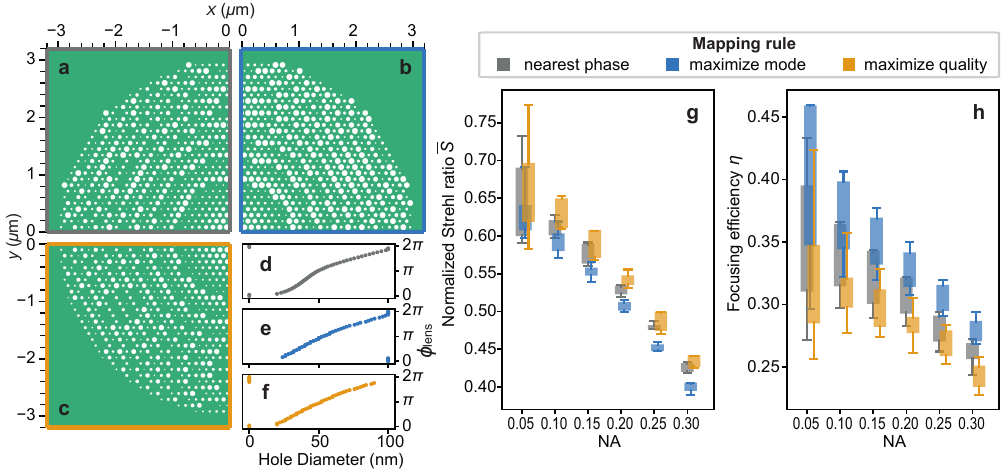}
\caption{Different meta-atom mapping rules with the hexagonal-lattice layout. \nolinebreak{Metalenses} with NA = 0.3 and $\Phi_g = 0.2\pi$ were obtained with mapping rules "nearest phase", "maximize mode", and "maximize quality", with their layouts presented in (a--c), respectively. (d--f) presents the mapping from phase to hole diameter for each mapping rule, respectively. The right panel summarizes the performance of different mapping rules for the hexagonal-lattice layout at different NAs, with the normalized Strehl ratio $\bar{S}$ shown in (g) and the focusing efficiency $\eta$ in (h). The definitions of the boxes and whiskers follow Fig.~\ref{fig:layout}.}
\label{fig:mapping_hex}
\end{figure}

\begin{figure}[htbp]
\centering
\includegraphics[scale=0.8,page=9]{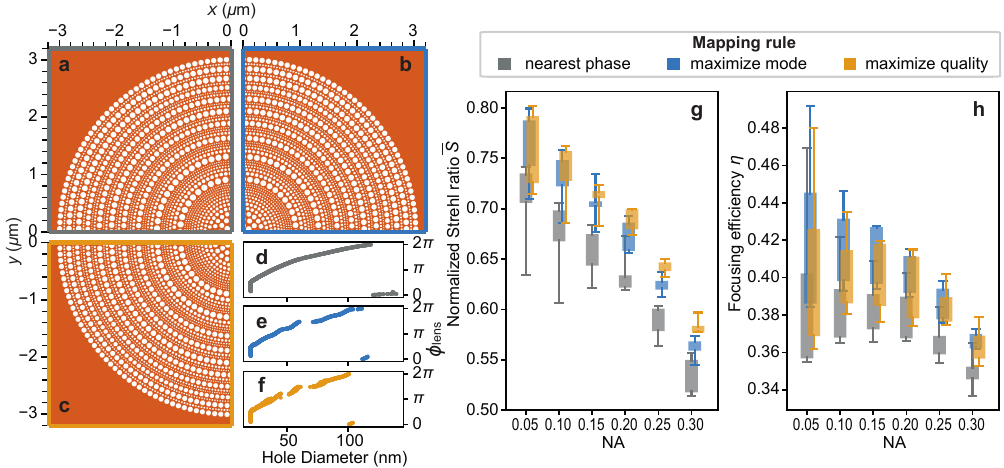}
\caption{Different meta-atom mapping rules designed with the fixed-gap ring layout scheme. Metalenses with NA = 0.3 and $\Phi_g = 0.2\pi$ were obtained with mapping rules "nearest phase", "maximize mode", and "maximize quality", with their layouts presented in (a--c), respectively. (d--f) presents the mapping from phase to hole diameter for each mapping rule, respectively. The right panel summarizes the performance of different mapping rules for the fixed-gap ring layout at different NAs, with the normalized Strehl ratio $\bar{S}$ shown in (g) and the focusing efficiency $\eta$ in (h). The definitions of the boxes and whiskers follow Fig.~\ref{fig:layout}.}
\label{fig:mapping_e2e}
\end{figure}

\begin{figure}[htbp]
\centering
\includegraphics[scale=0.8,page=10]{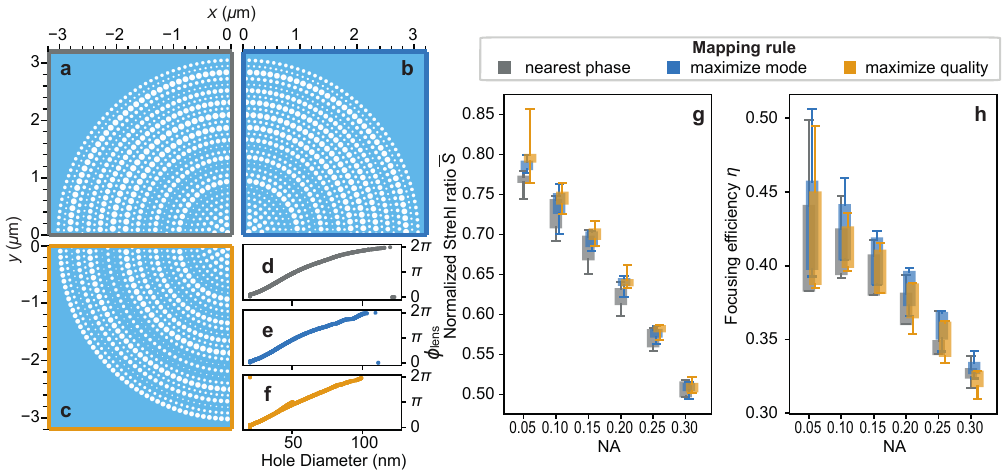}
\caption{Different meta-atom mapping rules designed with the max-$t_0$ ring layout scheme. Metalenses with NA = 0.3 and $\Phi_g = 0.2\pi$ were obtained with mapping rules "nearest phase", "maximize mode", and "maximize quality", with their layouts presented in (a--c), respectively. (d--f) presents the mapping from phase to hole diameter for each mapping rule, respectively. The right panel summarizes the performance of different mapping rules for the max-$t_0$ ring layout scheme at different NAs, with the normalized Strehl ratio $\bar{S}$ shown in (g) and the focusing efficiency $\eta$ in (h). The definitions of the boxes and whiskers follow Fig.~\ref{fig:layout}.}
\label{fig:mapping_fp}
\end{figure}

Comparisons of different mapping rules for metalenses with hexagonal lattice, fixed-gap ring and max-$t_0$ ring layout schemes are provided in Figs.~\ref{fig:mapping_hex}, \ref{fig:mapping_e2e} and \ref{fig:mapping_fp}, respectively. Similar to when the different mapping rules are applied in the design of metalenses using the square-lattice layout scheme, the major difference between the mapping of the different mapping rules for the case of the hexagonal-lattice layout scheme is at the phase gap of the metaatom library, i.e. near 0 and $2\pi$ phases: The "maximize mode" rule prefers large holes for their high zeroth order transmittances, whereas the "maximize quality" rule prefers absence of holes to minimize the transmittances of the higher diffraction orders. The lens performance with the hexagonal-lattice layout scheme shows a similar dependence on the mapping rules to the square-lattice layout scheme (Fig.~\ref{fig:mapping}). The "maximize quality" rule always yields the hexagonal-lattice metalens with the highest normalized Strehl ratio $\bar{S}$, improving the highest performance by 2\%--6\% across different NAs compared to the "nearest phase" rule. On the other hand, the "maximize mode" rule always provides the hexagonal-lattice metalens with the highest focusing efficiency $\eta$, improving the highest $\eta$ by 6\%--11\% across different NAs compared to the "nearest phase" rule.

For ring-type layouts, the mapping from phase to hole diameter is not one-on-one (Figs.~\ref{fig:mapping_e2e}d--\ref{fig:mapping_e2e}f, \ref{fig:mapping_fp}d--\ref{fig:mapping_fp}f) because the selection of a metaatom for a ring-type layout has to satisfy also Eqs.~(\ref{eq:interring}) and (\ref{eq:intraring}) in the  text. In both the fixed-gap and max-$t_0$ schemes, the "maximize mode" rule avoids selecting the largest holes in the metaatom library, limiting the maximum hole diameter to 115 nm and 102 nm, respectively. The "maximize quality" rule also avoids the largest holes, with even lower hole-diameter ceilings of 102 nm and 98 nm, respectively. 

For ring-type layout schemes, the "maximize quality" rule generally yields metalenses with the best $\bar{S}$. The metalens with the highest Strehl ratio for each NA are 6.8\% and 3.1\% higher for the fixed-gap scheme and the max-$t_0$ scheme, respectively, with the "maximize quality" mapping rule than with the "nearest phase" mapping rule. The "maximize mode" rule yields metalenses with the highest $\eta$ across all NAs using the max-$t_0$ ring layout scheme. Interestingly, for the fixed-gap ring layout scheme, while the "maximize mode" rule still provides the metalenses with the highest focusing efficiencies at low NA (0.05 to 0.20), the metalenses with the highest $\eta$ at higher NA (0.25 and 0.30) are from the "maximize quality" rule. The distinct behavior might be due to denser hole packing resulting from the combination of the fixed-gap layout scheme and the "maximize quality" mapping rule, which reduces the impact of increasing NA on aliasing and phase error.

It was previously observed with lattice-type layout schemes that the metalenses obtained from the "nearest phase" rule exhibit performance between those from the "maximize mode" and "maximize quality" rules, as measured by $\bar{S}$ and $\eta$. However, the performances of metalenses obtained from the "nearest phase" rule are not intermediate, but are inferior to the metalenses obtained from the "maximize mode" and "maximize quality" rules in general in terms of both figures of merit for the fixed-gap layout scheme and in terms of $\bar{S}$ for the max-$t_0$ layout scheme. The intermediate performance of the metalenses from the "nearest phase" mapping rule is likely a result from the fact that the mapping function obtained from the "nearest phase" rule is intermdiate between the mapping functions obtained from the "maximize mode" and "maximize quality" rules (Figs.~\ref{fig:mapping}d--\ref{fig:mapping}f, \ref{fig:mapping_hex}d--\ref{fig:mapping_hex}f). When the mapping function of the "nearest phase" rule is no longer the intermediate between the other two mapping rules, i.e., in the fixed-gap ring and max-$t_0$ ring layout schemes, the performance of the created metalenses no longer has a performance between metalenses created by the other two mapping rules. 

\section{Supplementary figures}

\begin{figure}[htbp]
\centering
\includegraphics[scale=0.8,page=4]{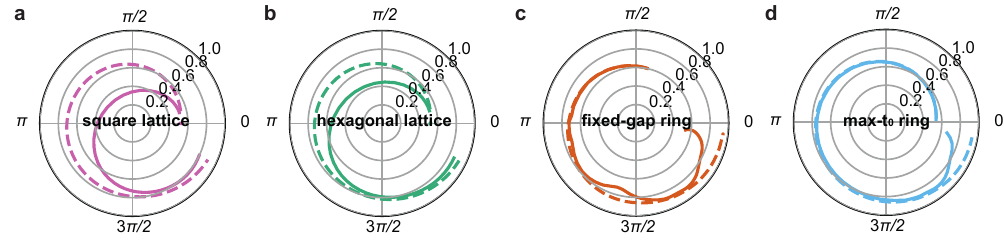}
\caption{The complex transmission coefficients of the zeroth diffraction order of the available metaatoms in (a) square lattice, (b) hexagonal lattice, (c) fixed-gap ring and (d) max-$t_0$ ring layout schemes are plotted in the complex plane with solid lines. The square root of the total transmission of the corresponding metaatoms is plotted on the same complex plane with a dashed line.}
\label{fig:metaatom_t_coeff}
\end{figure}

\begin{figure}[htbp]
\centering
\includegraphics[page=1]{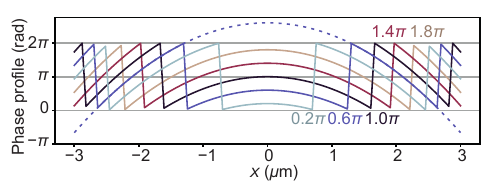}
\caption{Lens phase profiles, following Eq.~(\ref{eq:phaseprofile}) in the manuscript, crossing the center of a 6-$\mu$m-diameter EUV metalens with focal length 59.9 $\mu$m (corresponding to a lens NA of 0.05) for $\Phi_g=0.2\pi, 0.6\pi, 1.0\pi, 1.4\pi$ and $1.8\pi$. The value of $\Phi_g$ for each plot is labeled with the corresponding color. Additionally, an unwrapped version of the phase profile for $\Phi_g = 0.6\pi$ is presented with a dashed line.}
\label{fig:lens_phase_profile}
\end{figure}

\begin{figure}[htbp]
\centering
\includegraphics[page=2]{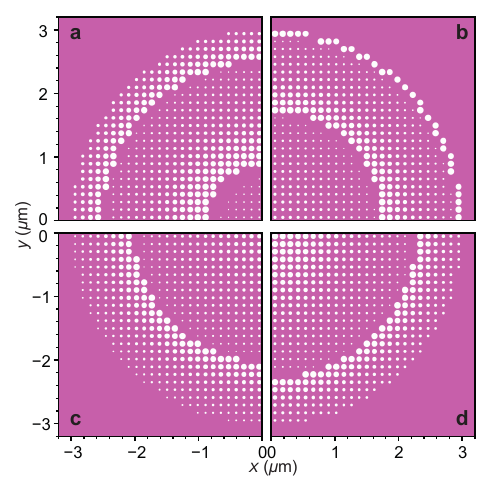}
\caption{Square-lattice metalens layouts of NA 0.05 are presented for global phase shift $\Phi_g$ (a) $0.2\pi$, (b) $1.0\pi$, (c) $1.4\pi$ and (d)$1.8\pi$. White color represents holes, whereas the pink colored region represents silicon. Together with Fig.~\ref{fig:layout}a ($\Phi_g=0.6\pi$) in the manuscript, the layouts of the five realizations of square-lattice metalens of NA 0.05, mapped with the "nearest phase" rule, are provided.}
\label{fig:square_layouts}
\end{figure}

\begin{figure}[htbp]
\centering
\includegraphics[scale=0.8,page=3]{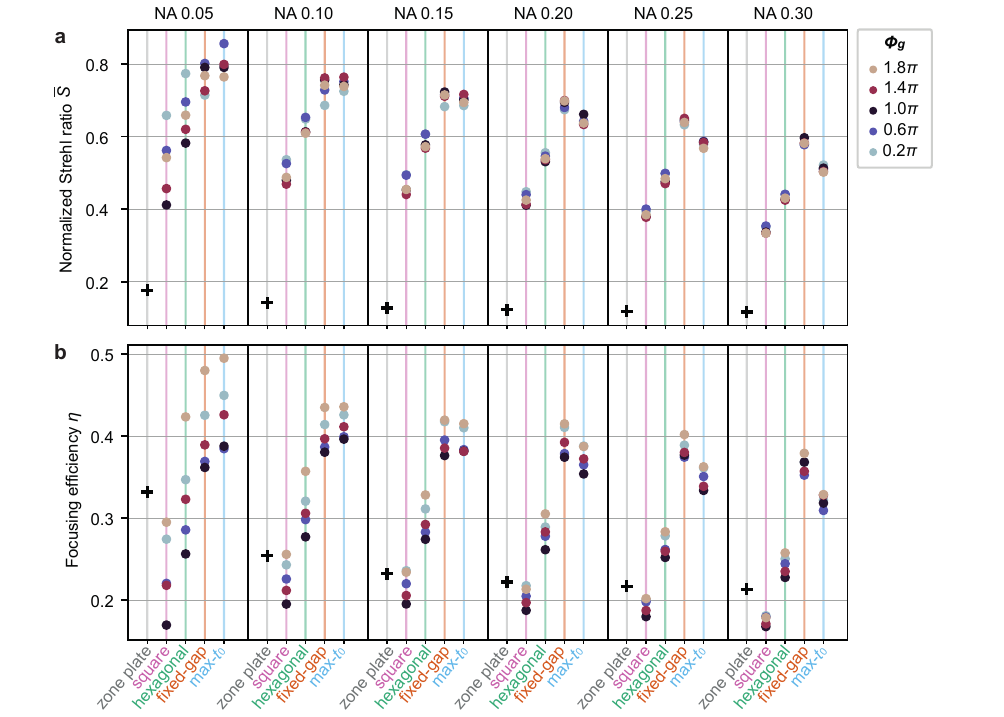}
\caption{The (a) normalized Strehl ratio $\bar{S}$ and (b) focusing efficiency $\eta$ of the binary zone plates and metalenses presented in Fig.~\ref{fig:layout}. The color of each marker represents the global phase $\Phi_g$ of the corresponding metalens. The metalenses were obtained with the "nearest phase" mapping rule.}
\label{fig:layout_performance_detailed}
\end{figure}

\begin{figure}[htbp]
\centering
\includegraphics[scale=0.8,page=12]{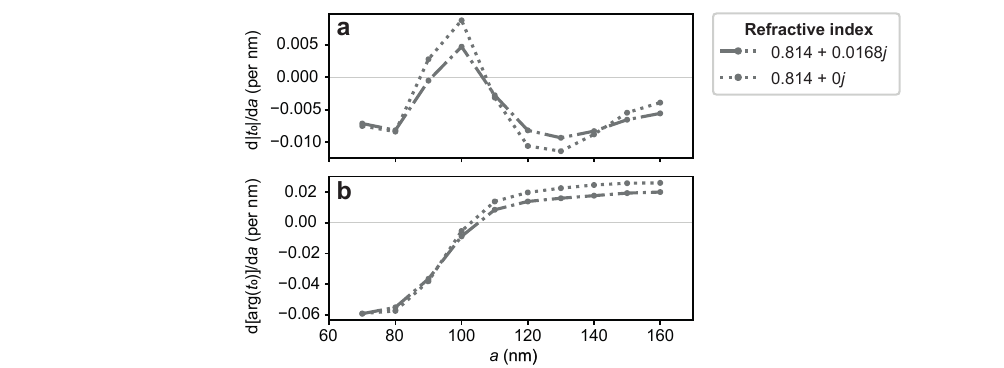}
\caption{The derivative of the (a) magnitude and (b) argument of the zeroth diffraction order transmission coefficient $t_0$ with respect to the lattice constant $a$ for hexagonal-lattice nanohole arrays with hole diameter 60 nm. The two curves in each subfigure correspond to the case with a membrane with the actual refractive index of crystalline silicon and a hypothetical case of lossless crystalline silicon. The data suggest that the sensitivity of $t_0$ of the crystalline silicon nanoholes to the separation distance from neighboring holes is reduced by absorption.}
\label{fig:derivative_t_a}
\end{figure}

\begin{figure}[htbp]
\centering
\includegraphics[scale=0.8,page=11]{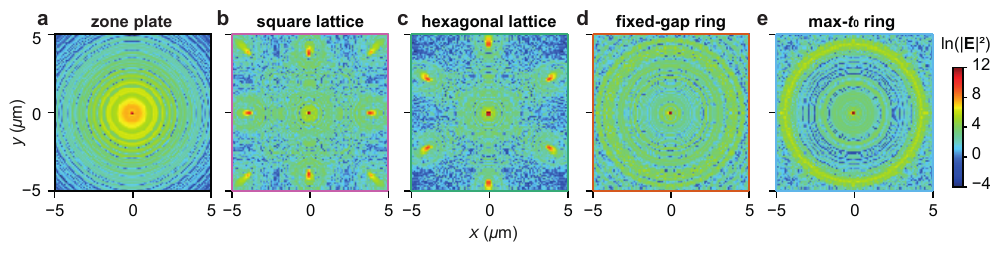}
\caption{Electric field intensity in logarithmic scale at the focal plane of (a) the binary zone plate, (b) the square-lattice layout metalens, (c) the hexagonal-lattice layout metalens, (d) the fixed-gap ring layout metalens, and (e) the max-$t_0$ ring layout metalens presented in Fig.~\ref{fig:prop}. The NA of all focusing elements is 0.30 (corresponds to a focal length of 9.54 $\mu$m). The global phase $\Phi_g$ of the metalenses is fixed at 1.8$\pi$. The focal plane is defined here as the plane 9.54 $\mu$m after the output plane of the focusing element.}
\label{fig:focalplane}
\end{figure}

\begin{figure}[htbp]
\centering
\includegraphics[scale=0.8,page=5]{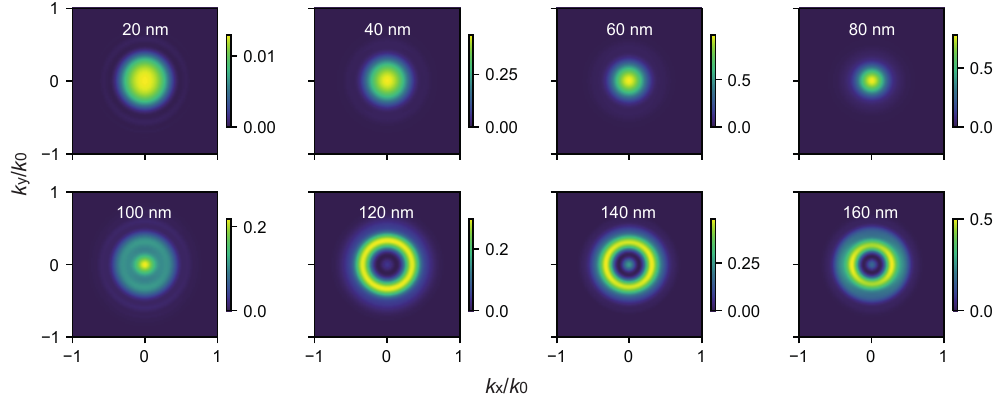}
\caption{The far fields of nanoholes of diameters 20 nm to 160 nm in a 280-nm-thick silicon membrane. The horizontal and vertical axes are the normalized transverse wavevectors. The hole diameters are labeled in the plots. The field intensity has an arbitrary unit.}
\label{fig:ff_2d}
\end{figure}

\begin{figure}[htbp]
\centering
\includegraphics[scale=0.8,page=6]{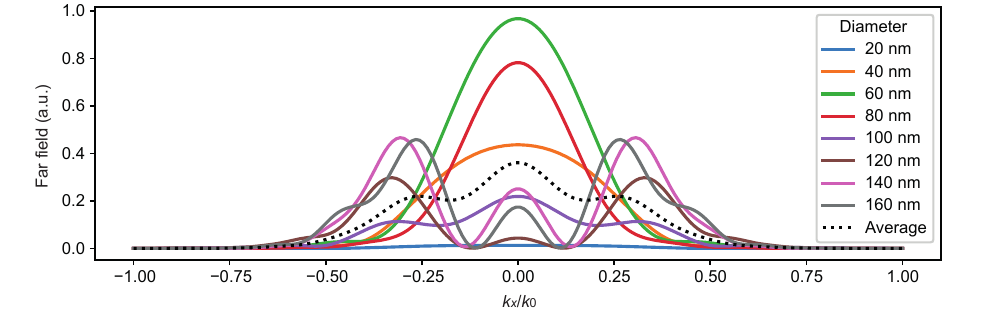}
\caption{Cross sections of Fig.~\ref{fig:ff_2d} at $k_y/k_0=0$. The dotted line represents the average of the far fields for all the hole diameters presented in the figure. The field intensity has an arbitrary unit.}
\label{fig:ff_1d}
\end{figure}

\begin{figure}[htbp]
\centering
\includegraphics[scale=0.8,page=7]{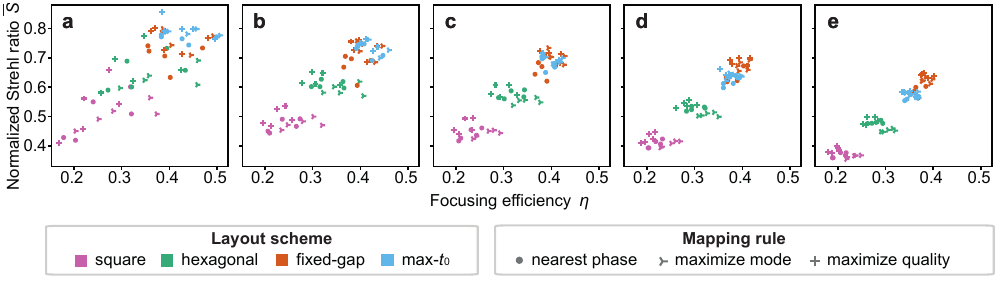}
\caption{Summaries of the performances of (a) NA 0.05, (b) NA 0.10, (c) NA 0.15, (d) NA 0.20 and (e) NA 0.25 EUV metalenses, designed with the different layout schemes (square, hexagonal, fixed-gap and max-$t_0$) and mapping rules ("nearest phase", "maximize mode" and "maximize quality") proposed in this paper. The data for NA 0.30 is presented in Fig.~\ref{fig:NA0.3}.}
\label{fig:summary_allNA}
\end{figure}

\end{document}